%% file: main.tex
\documentclass[qubs,article,accept,moreauthors,pdftex,10pt,a4paper]{mdpi} 
\setitemize{parsep=6pt,itemsep=0pt,leftmargin=*,labelsep=5.5mm}
\setenumerate{parsep=6pt,itemsep=0pt,leftmargin=*,labelsep=5.5mm}
\setlist[description]{itemsep=0mm}   
\usepackage{upgreek,color}
\usepackage{gensymb}
\usepackage{tikz}
\usetikzlibrary{decorations.pathreplacing}
\usetikzlibrary{decorations.markings}
\usepackage[color=red!70, loadshadowlibrary, shadow, colorinlistoftodos]{todonotes}

\usepackage{subcaption}
\usepackage{algorithm2e}
\SetKwComment{Comment}{/* }{ */}
\definecolor{neutron}{HTML}{0047AB}
\firstpage{1} 
\makeatletter 
\setitemize{parsep=6pt,itemsep=0pt,leftmargin=*,labelsep=5.5mm}
\setenumerate{parsep=6pt,itemsep=0pt,leftmargin=*,labelsep=5.5mm}
\setlist[description]{itemsep=0mm}   
\usepackage{upgreek,color}
\firstpage{1} 
\pubvolume{1}
\issuenum{1}
\articlenumber{1}
\pubyear{2025}
\copyrightyear{2025}
\history{Received: 2025}
\updates{yes} 

\usepackage{placeins}

\Title{Simulating neutron diffraction from deformed mosaic crystals in McStas }

\Author{Daniel Lomholt Christensen$^{1}$\orcidA{}, Sandra Cabeza$^{2}$\orcidD{}, Thilo Pirling$^{2}$\orcidE{}, Kim Lefmann$^{1}$\orcidB{}, Jan Saroun$^{3}$\orcidC}
\AuthorNames{Daniel Lomholt Christensen and Jan Saroun}

\address{%
$^{1}$ \quad Nanoscience Center, Niels Bohr Institute, University of Copenhagen, DK-2100 Copenhagen {\O} , Denmark\\
$^{2}$ \quad Institut Laue-Langevin, F-38000 Grenoble, France\\ 
$^{2}$ \quad Nuclear Physics Institute CAS, Rez 25068, Czech Republic\\
}

\abstract{
Monochromator and analyzer systems that rely on bent single crystals are in use throughout the neutron scattering community. We here introduce a new component to the neutron simulation software package McStas, that simulates these bent single crystals. We then compare the performance of this component to like software in SIMRES, and Ncrystal. These simulations show excellent agreement across the different software programs. Finally we compare simulations of the new McStas component with analytical calculations of secondary extinction, based on the Darwin transfer equations. Here we also find excellent agreement, further validating this new component. 
}


\begin{document}

\section{Introduction}
\label{sect:intro}
The scientific request for neutron scattering experiments is vast, as can be seen from the building and maintenance of large scale facilities such as the ILL, ESS, ISIS, J-PARC, CSNS, SNS etc.\cite{noauthor_ess_nodate, noauthor_isis_nodate, noauthor_ill_nodate, noauthor_china_nodate, 
noauthor_j-parcjapan_nodate,
noauthor_spallation_nodate}. Every new instrument constructed or upgraded at such large facilities inevitably involves a large cost \cite{ILLMILLENIUM, ILLENDURANCE}. Simulating instruments, and upgrades to instruments, before they are built allows for estimation of potential instrument efficiency and resolution. This leads to the possibility of better informed decisions, therefore improving both economy and scientific performance of the facilities.

An essential ingredient of many neutron scattering instruments is Bragg diffraction off single crystals, used for both monochromators or analyzers. Most often, these crystals are imperfect due to the mosaicity - the presence of randomly misoriented perfect crystal domains. Such a mosaicity is usually desired to increase the integral reflectivity of the crystal. Alternatively, the reflectivity can also be increased by elastic bending of a perfect single crystal. This technique has been employed in the construction of instruments which require high resolution combined with focusing effects, such as diffractometers measuring residual strains \cite{SALSA,STRESSPEC,KOWARI} or neutron spectrometers \cite{THALES}. 
Recent development of plastically bent perfect Si crystals as multianalyzers for neutron spectrometers \cite{MARMOT} underlines the need for fast, yet realistic algorithms for simulation of neutron transport in bent crystals.

McStas \cite{McStas1, McStas2, McStas3} is a software package used worldwide for simulating neutron instruments by using the neutron ray tracing method. McStas includes a vast library of models for various components used in existing and newly designed neutron instruments. A model describing undeformed mosaic crystals has been included since the beginning of McStas development, and very realistic simulations of such crystals have recently been enabled by the binding of McStas with the NCrystal package \cite{cai_ncrystal_2020,kittelmann_elastic_2021}. However, an efficient model for bent perfect or mosaic crystals is currently not supported.

An algorithm for simulation of the bent perfect and mosaic crystals has previously been developed by one of us \cite{NIMA} for the software package SIMRES \cite{saroun_restrax_1997, simres_github}. In this work, we port this algorithm from SIMRES to McStas by creating a new component named \textit{Monochromator\_bent.comp}. The underlying theory is described in the next section, followed by the simulation results. The new component is validated by comparison of simulations for identical setups using both packages and by comparing the simulated reflectivity with analytical calculations for the limiting case of a flat mosaic crystal.  


\section{Reflectivity of deformed crystals}
\FloatBarrier
\label{sect:reflectivity}

\begin{figure*}
    \centering
    \begin{subfigure}{0.49\textwidth}
        \centering
        \resizebox{\textwidth}{!}{
        \input{Figures/Unbent_crystal}
        }
        \label{fig:unbent_crystal}
    \end{subfigure}
    ~ 
    \begin{subfigure}{0.49\textwidth}
        \centering
        \resizebox{\textwidth}{!}{
        \input{Figures/Single_crystal.tex}
        }
        \label{fig:deformed_crystal}
    \end{subfigure}
    ~
    \caption{(Left) An unbent crystal. The reflecting lattice planes are cut at an angle $\chi$ to the surface. (Right) The deformed crystal with $\chi=0$. The reciprocal lattice vectors are shown as $\boldsymbol{\tau}_0$ and $\boldsymbol{\tau}(\boldsymbol{r})$ the unbent, and bent reciprocal lattice vector, respectively.}
    \label{fig:deformed_crystal_fig}
\end{figure*}
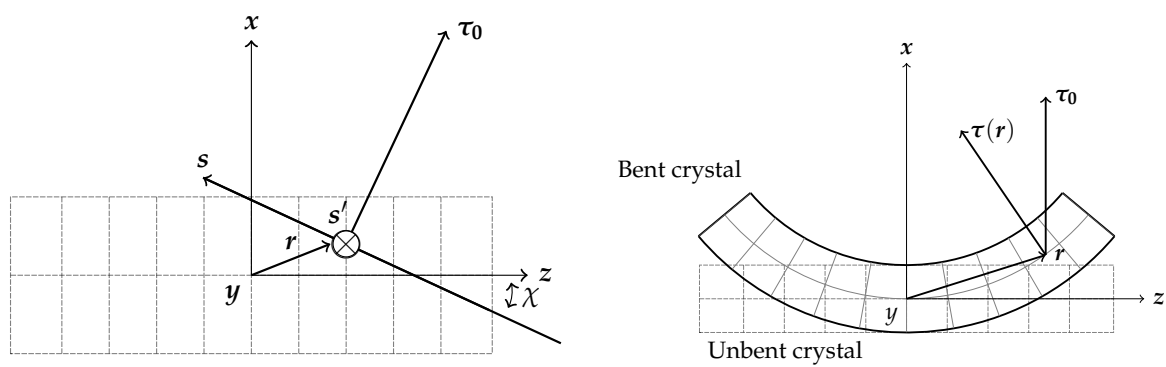
\label{sec:Bragg scattering in bent mosaic crystals}
The neutron ray tracing algorithm for bent crystals, introduced in \cite{NIMA}, is based on the theory first introduced by \citet{Hu}, which yields analytical expressions for the reflecting power of bent mosaic crystals. In this section, we follow the derivation in \cite{NIMA}, but with more detail and a few additions. 
 
The algorithm employs several approximations: (a) The orientation of single crystal domains (mosaic blocks) are uncorrelated as assumed for the type I mosaic single crystals described by \citet{ZACHARIASEN}. (b) Reflection by each mosaic block is treated kinematically as for a thin plane parallel single crystal, but a correction for primary extinction can be added as a constant multiplication factor to the kinematic reflectivity. (c) The two-beam approximation is adopted, assuming only the direct beam and the beam reflected by a single lattice plane defined by a selected reciprocal lattice vector. (d) The crystal bending (if applied) is elastic and uniform, producing a constant gradient of deviation of the reciprocal lattice vector from its nominal (undeformed) value. (e) Quasi-classical trajectories of neutrons in uniformly bent perfect crystals are hyperbolic \cite{Lukas}. However, we assume that the bending is strong enough so that the trajectories can be approximated by straight line segments with singular points of reflection. 


These approximations are typically justified for monochromators and analyzers, where only neutron flux and instrument resolution are of main interest. For other applications, the model may be still too idealistic, ignoring some important features of real crystals, such as the presence of other competing reflecting planes, correlated mosaicity, or dynamical diffraction effects in the case of weakly bent perfect crystals. The component should thus be used with these limitations in mind. 

\subsection{Scattering power} \label{sec_scattering_power}

The notation and reference frames used in the following sections are shown in Figure \ref{fig:deformed_crystal_fig}. Namely, $\boldsymbol{\tau}_0$ is the mean reciprocal lattice vector of the undeformed crystal, $\mathbf{s}$ is a unit vector normal to $\boldsymbol{\tau}_0$ in the horizontal plane, $\mathbf{s}'$ is the unit vector along the y axis and $\chi$ is the crystal cutting angle with respect to the reflection plane.

The key parameter for neutron transport calculations is the scattering power, i.e. the probability of reflection per unit path length for a thin crystal layer with negligible extinction. It can be expressed as the macroscopic scattering cross section per unit volume,
\begin{equation}
    \Sigma(\mathbf{r}) = \frac{Q}{\eta} g\left(\frac{\epsilon(\mathbf{r})}{\eta}\right),
    \label{eq:Cross_section}
\end{equation}
where $Q$ is the kinematic reflectivity, 
$Q=|F|^2\lambda^3/(\sin(2\theta)V_0^2)$ multiplied by the Debye Waller factor. $g$ is the distribution function for mosaic domain angles in the $z,x$ plane of unit width and $\eta$ is the width of this distribution. In the expression for $Q$, $\lambda$ is the neutron wavelength, $\theta$ is the Bragg angle, $F$ is the scattering amplitude of the unit cell and $V_0$ is the unit cell volume. The variable $\epsilon$ is the angular deviation of a mosaic domain required to satisfy the Bragg condition for a given incident wave vector. In a deformed crystal, $\epsilon$ becomes a function of position, $\mathbf{r}$. Using Equation \ref{eq:Cross_section}, the transmission probability of a neutron through a crystal is then found by the Beer-Lambert law,
\begin{equation}
	P_T = \exp{\left(-\int_{L}\Sigma(\mathbf{r})\text{d}\mathbf{l}\right)},
	\label{eq:Lambert_Beers}
\end{equation}
where the integral is a line integral over the neutron path through the crystal.

In order to find $\epsilon(\mathbf{r})$, one begins with the Bragg condition in vector form
\begin{equation}
    \mathbf{k}_i - \mathbf{k}_f = \boldsymbol{\tau}(\mathbf{r}) ,
    \label{eq:Bragg_condition_plain}
\end{equation}
where $\mathbf{k}_i (\mathbf{k}_f)$ is the incoming (outgoing) wavevector and $\boldsymbol{\tau}(\mathbf{r})$ is the spatially dependent reciprocal lattice vector for the selected Bragg reflection.
Using that the scattering is elastic, a rearrangement transforms this equation into    
\begin{equation}
    |\mathbf{k}_i + \boldsymbol{\tau}(\mathbf{r})|^2 - |\mathbf{k}_i|^2 = 0 .
    \label{eq:Bragg_condition}
\end{equation}
 The assumption of uniform deformation implies that $\boldsymbol{\tau}$ is a linear function of position,
\begin{align}
	\boldsymbol{\tau}(\mathbf{r}) &=\boldsymbol{\tau}_0 + \boldsymbol{\nabla\tau}\cdot \mathbf{r} - 	  
	|\boldsymbol{\tau}_0| \left( \epsilon \mathbf{s} + \gamma \mathbf{s}'  \right),
	\label{eq:tau_r}
\end{align}
where $\boldsymbol{\nabla\tau}$ is the gradient of $\boldsymbol{\tau}(\mathbf{r})$ derived in Appendix \ref{appendix: deformation gradient}. The third term in Equation \ref{eq:tau_r} is understood as the deviation due to mosaicity, assuming small, random misfit angles $\epsilon$ and $\gamma$ around the $\mathbf{s}'$ and $\mathbf{s}$ axes, respectively. Equation \ref{eq:tau_r} thus defines the diffraction vector in a bent mosaic single crystal. In a practical implementation, the diffraction vector has to be normalized to $|\boldsymbol{\tau}_0 + \boldsymbol{\nabla\tau}\cdot \mathbf{r}|$ as the adopted small-angle approximation slightly changes the length of the vector.

The function $\epsilon(\mathbf{r})$ is then found by substituting Equation \ref{eq:tau_r} into Equation \ref{eq:Bragg_condition},
\begin{align}
	\left| \mathbf{k}_i + \boldsymbol{\tau}_0 + \boldsymbol{\nabla\tau}\cdot \mathbf{r} - |\boldsymbol{\tau}_0|\left( \epsilon \mathbf{s} + \gamma \mathbf{s}'  \right) \right|^2 - 
	|\mathbf{k}_i|^2 &=0.
	\label{eq:Bragg_law_full}
\end{align}
In the expansion of Equation \ref{eq:Bragg_law_full}, we make following approximations: The bending is small, meaning that second order terms of  $\boldsymbol{\nabla\tau}\cdot \mathbf{r}$ are neglected. This is justified in practice when the bending radius is always much larger than the crystal dimensions. Similarly, we neglect second order terms for $\epsilon$ and $\gamma$ since the mosaicity is typically smaller than $1^\circ$. Finally, we assume that  $\boldsymbol{\tau}_0$ and $\mathbf{k}_i$ are both in (or near) the $z,x$ plane, which is also typical for crystals used as monochromators or analyzers.
Equation \ref{eq:Bragg_law_full} then yields $\epsilon(\mathbf{r})$  in the first order approximation as,

\begin{align}
	\epsilon(\mathbf{r}) &= \frac{(\mathbf{k}_i + \boldsymbol{\tau}_0 )^2 - |\mathbf{k}_i|^2}{2|\boldsymbol{\tau}_0| \mathbf{k}_i \cdot \mathbf{s}} + 
	\frac{(\mathbf{k}_i + \boldsymbol{\tau}_0) \cdot \nabla\boldsymbol{\tau} \cdot \mathbf{r} } {|\boldsymbol{\tau}_0| \mathbf{k}_i \cdot \mathbf{s}} 
	 \label{eq:eps_r}
\end{align}
The incident neutron trajectory has the linear form $\mathbf{r} = \mathbf{r}_0 + \mathbf{v}_i t$, where $\mathbf{v}_i$ is the incident neutron velocity and $t$ is the flight time. 
Equation \ref{eq:eps_r} remains valid for any coordinate origin, $\mathbf{r}_0$ if $\boldsymbol{\tau}_0$ is replaced with $\boldsymbol{\tau}_0 + \boldsymbol{\nabla\tau}\cdot \mathbf{r}_0$. We can thus substitute the position of the incident neutron in Equation \ref{eq:eps_r} and arrive to $\epsilon$ expressed as the function of flight time,
\begin{align}	  
	\epsilon(t) &=  \epsilon_0 + \beta t
	\label{eq:eps_t}
\end{align}
\begin{align}
	\text{where} \quad
	\epsilon_0 \equiv \frac{(\mathbf{k}_i + \boldsymbol{\tau}_0 )^2 - |\mathbf{k}_i|^2}{2|\boldsymbol{\tau}_0| \mathbf{k}_i \cdot \mathbf{s}} \quad,\quad
	\beta \equiv \frac{(\mathbf{k}_i + \boldsymbol{\tau}_0) \cdot \nabla\boldsymbol{\tau} \cdot \mathbf{v}_i }{|\boldsymbol{\tau}_0| \mathbf{k}_i \cdot \mathbf{s}} 	  
	\label{eq:eps0_beta}
\end{align}
for $\boldsymbol{\tau}_0$ evaluated at the flight origin, $\mathbf{r}_0$.

The equations (\ref{eq:Cross_section}), (\ref{eq:Lambert_Beers}) and (\ref{eq:eps_t}) allow us to express the cumulative probability of reflection as a function of flight time as
\begin{equation} 
	P(t) =  1 - P_T = 	
	1 - \exp\left(-\frac{Q|\mathbf{v}_i|}{\beta}\left[\Phi\left(\frac{\epsilon_0+\beta t }{\eta}\right)-\Phi\left(\frac{\epsilon_0}{\eta}\right)\right]\right),
	\label{eq:Diff_final}
\end{equation}
where $\Phi$ is the cumulative probability for the distribution $g$. The McStas implementation assumes that $g$ is the Gaussian distribution and $\Phi$ thus equals to $(1+erf)/2$. However, this algorithm works also for any other form of mosaicity distribution.

For a \textit{bent perfect crystal}, Equation \ref{eq:Diff_final} must be evaluated in the limit $\eta \rightarrow 0$. It yields non-zero $P(t)$ only if $\epsilon_0+\beta t = 0$ for some positive $t < t_{max}$, where $t_{max}$ is the flight time to the crystal exit, i.e. when the Bragg condition is met on the path through the crystal. Then the probability 
\begin{equation} 
P_0(t_{max}) = 1 - \exp\left(-\frac{Q |\mathbf{v}_i|}{|\beta|}\right) 
\end{equation}
determines the crystal reflectivity.

\subsection{Finding the point of scattering}

The scattering position can be found by sampling of $t$ from the cumulative probability distribution $P$ in Equation \ref{eq:Diff_final}. This is done with the inverse transformation sampling \cite{devroye_non-uniform_2014}. Provided that only the reflected neutrons are of interest, we can normalize the distribution by dividing with the transmission probability, $P(t_{max})$. In this way, no neutron history would be lost, but neutron weight has to by multiplied by $P(t_{max})$. Alternatively, it is possible to set this normalization factor $P(t_{max})=1$ and allow for transmission if $t > t_{max}$. To obtain $t$ sampled with the distribution from Equation \ref{eq:Diff_final}, we replace $\frac{P(t)}{P(t_{max})}$ with a uniform random variable  $\xi$  distributed between 0 and 1 and isolate $t$,
\begin{align}
    t &= \frac{\eta}{\beta}\Phi^{-1}\left(\Phi\left(\frac{\epsilon_0}{\eta}\right)-\frac{\beta}{Q|\mathbf{v}_i|}\ln(1-\xi P(t_{max}))\right)-\frac{\epsilon_0}{\beta}
    \label{eq:t_max}.
\end{align}
In the limit $\eta \rightarrow 0$, the above reduces to $t = -\epsilon_0/\beta$, which is the flight time to the position where the neutron  meets the Bragg condition. This equation is only true, provided that this position is inside the crystal. The transport through a bent perfect crystal thus becomes deterministic. 

\begin{figure}
    \centering
    \includegraphics[width=0.6\textwidth]{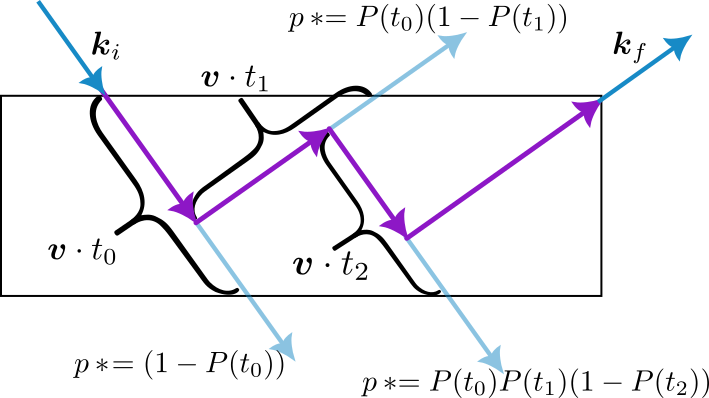}
    \caption{A neutron ray (purple) tracing a possible path through the crystal. $t_n$ is the n'th $t_{max}$.Notice that the algorithm uses $t_{max}$ to sample a $t$ value at every reflection. The figure is heavily inspired by \cite{NIMA}.}
    \label{fig:Tracing_a_ray}
\end{figure}

With these tools in hand, we can now simulate multiple scattering within the crystal, since we can just repeat scattering as required. Such a multiple scattering event is shown in Figure \ref{fig:Tracing_a_ray}, where a neutron ray goes through three possible events, having the choice of reflecting, or transmitting at each sampled time $t$. 

In the component this multiple reflection pattern is performed across all crystals the user defines in an array. This allows for not only having multiple scattering inside of one crystal, but scattering in multiple crystals, in a back and forth manner.

\subsection{Corrections for absorption, scattering and primary extinction}
Since the total distance traveled through the crystal, $l$ is available, beam attenuation by absorption and other scattering events can be accounted for by multiplying each neutron weight by the factor:
\begin{align}
    A = \exp\left(-\mu_{tot}l\right),
    \label{eq: absorption}
\end{align}
where $\mu_{tot}$ is the linear attenuation coefficient including the effects of absorption, incoherent scattering and phonon scattering.
The McStas component uses the calculated values of $\mu_{tot}$ according to \citet{freund_cross-sections_1983} for selected crystal materials.

The primary extinction is accounted for by multiplying the kinematic reflectivity with the theoretical extinction function derived for a crystal slab in symmetric Bragg geometry \cite{ZACHARIASEN},
\begin{align}
    \Upsilon &= \frac{\tanh(d/l_e)}{d/l_e}										
\end{align}
where $l_e = V_0/(\lambda F)$ is the extinction length and $d$ is the flight path through the slab. 
Since the actual shape of mosaic blocks is not known, $d$ should be interpreted as an effective domain size, to be determined for example by fitting of the simulated reflectivity to experimental data. Typically, values of the order of 10 $\mu$m should describe well the observed reflectivity in mosaic crystal monochromators. 

\subsection{Algorithm}
\label{sect:algorithm}
The implementation of the bent mosaic crystal rests on quite a simple algorithm. It is written out as pseudo code in Algorithm \ref{alg:Crystal}, for multiple crystals in an array. The general idea is to check if the neutron ray hits the crystal array (a), if yes, find the probabilities of reflection for all crystals  (b), choose which crystal to reflect from (c), sample the reflection time and propagate to the point of reflection (d) and then reflect the neutron ray (e). While the neutron ray is reflected, the weight $p$ is updated to reflect any Monte  Carlo choices (f). This process is repeated, allowing for multiple reflections. The distance $L$ traveled inside the crystal is summed and the weight $p$ is then multiplied by the attenuation factor $A$ from Equation \ref{eq: absorption}. 

\begin{algorithm}
    \caption{Pseudo-code explaining the algorithm behind \textit{Monochromator\_Bent}. The comments in parenthesis refer to the descriptions in the text.}
    \label{alg:Crystal}
    find\_intersections(monochromator, neutron)\ \Comment*[r]{(a)}
    \If{$intersections = \emptyset$}{
        Allow neutron to pass on without interaction\;
    }
    \While{infinite loop}{
        Find\_the\_probability\_of\_reflection\_for\_all\_intersected\_crystals\ \Comment*[r]{(b)}

        Choose\_crystal\_to\_reflect\_on\ \Comment*[r]{(c)}

        \If{Neutron does not reflect}{
            break;
        }

        Sample\_point\_of\_reflection\ \Comment*[r]{(d)}

        Propagate\_to\_point\_of\_reflection\ 

        Reflect\_neutron\ \Comment*[r]{(e)}
        update\_$p$\ \Comment*[r]{(f)}
        
        Find\_\_new\_intersections\;
    }
    $p *= A$ \Comment*{(g)}
\end{algorithm}


\section{Simulation setup}
\label{sect:setup}

\begin{figure}
    \centering
    \input{Figures/instrument_setup} 
    \caption{The setup used for simulating the bent mosaic crystal in both SIMRES and McStas. $x,y$ and $z$ is the coordinate system of the monochromator, and $\omega$ is the rotation angle around the $y$ axis, used for the rocking curve.}
    \label{fig:Instrument_setup}
\end{figure}
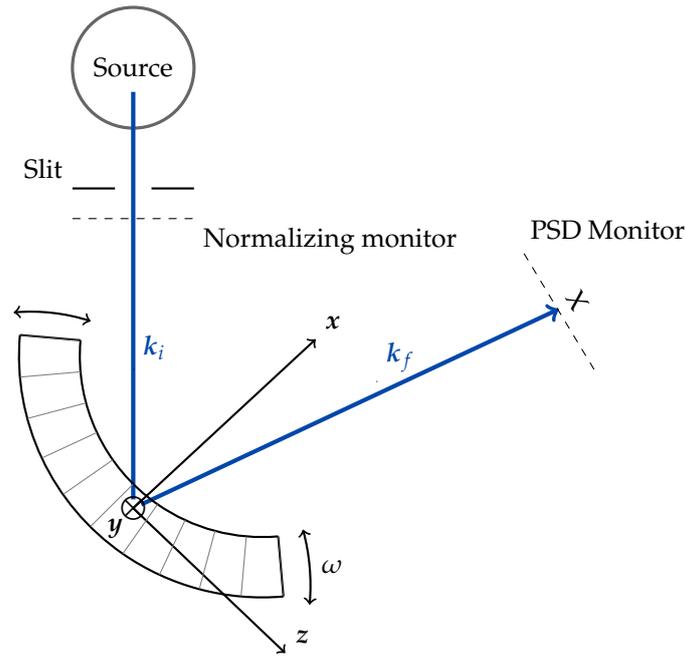

To validate the new crystal component for McStas, simulations of simple instruments were performed using both SIMRES and McStas. The general setup can be seen in figure \ref{fig:Instrument_setup}. 

A narrow monochromatic incident beam with very small divergence was used to probe the crystal reflectivity. The incident beam was a 0.1$\times$1~cm$^2$ (width $\times$ height) source that generated neutrons with a wavelength of $\lambda = 1.5\pm 1.5\cdot10^{-5}$\AA, where the position and wavelength were uniformly distributed. The neutrons were focused by a Monte-Carlo choice, to be sent through a 0.1$\times$1~cm$^2$ slit placed 10~m after the source. At the slit, a PSD monitor was also placed, such that normalization was possible.

A Germanium crystal\footnote{Ge(511): Structure factor  $|F|^2=0.33$~barns, unit cell volume $V_0=22.63$~\AA, lattice spacing $d=1.09$~\AA. For more details see the simulation and component code at \cite{christensen_simulations_2025}.} with dimensions 0.8$\times$1.2$\times$7~cm$^3$ (thickness $\times$ height $\times$ length) was then placed 0.15~m after the slit. The $y,z$ cutting surface was parallel to the (211) plane, but the (511) reflection (Bragg angle $\theta_{511} = 43.53^\circ$) was chosen for simulation, which resulted in an asymmetric reflection geometry with the diffraction vector rotated by $\chi = 19.47^\circ$ with respect to the $y,z$ surface. The crystal was thus rotated by $\theta_{511} - \chi = 24.06^\circ$ to meet the Bragg condition. This asymmetry was used in order to showcase the correctness of the $\chi$ variations and to better visualize the dependence of scattering power on the penetration depth and crystal curvature. 

Finally, a PSD monitor is placed 0.05~m after the crystal, at $2\theta_{511}$ to observe the Bragg scattering.

This setup is used to simulate three cases: (i) the bent perfect crystal for bending radii $R$ of 3, 5 and 10~m with $\eta=0'$, (ii) the bent mosaic crystal for the same radii and  $\eta=6'$ and (iii) a flat mosaic crystal by setting $R = 1000$ m and  $\eta=6'$. The case of $\eta=0'$ and $R = 1000$~m is not simulated, because the model cannot describe diffraction correctly near the limit of undeformed perfect single crystals. 

Further simulations comparing the new component to the Darwin transfer equations, as described in \cite{lefmann_neutron_1997}, were performed to model the effect of secondary extinction. Here, case (iii) with $6'$ mosaicity and $R=1000$ m is simulated in two different scans. The first scan varies the thickness of the crystal gradually while the reflectivity of the beam is monitored. The second scan is a rocking curve, varying $\omega$ around the Bragg angle, whilst recording the reflectivity at each $\omega$. The reflectivity in the Bragg case is,
\begin{align}
\begin{aligned}
    R_B &= \frac{2\alpha\frac{Q\Upsilon}{\eta}}{\left(\alpha + \varphi\right)\left(\frac{Q\Upsilon}{\eta}+ \mu_{tot}\right) + \gamma_B\coth\left(\frac{D\gamma_B}{2}\right)}\\
    \gamma_B &= \sqrt{\left(\alpha+ \varphi\right)^2\left(\frac{Q\Upsilon}{\eta} + \mu_{tot}\right)^2 - 4\alpha\varphi\left(\frac{Q\Upsilon}{\eta}\right)^2}.
    \label{eq:Analytical_mosaic_reflectivity}
\end{aligned}    
\end{align}
Here $R_B$ is the reflectivity, $D$ the thickness of the sample, $\alpha=\sec(\theta_{511} + \chi)$ and $\varphi = \sec(\theta_{511}-\chi)$ and all the other symbols are defined previously. For the simulations that compare to the Darwin transfer equations, we simulate also the copper (311)\footnote{Cu(311): Structure factor $|F|^2=0.60$~barns, unit cell colume $V_0=11.81$~\AA, lattice spacing $d=1.09$~\AA. For more details see the simulation and component code at \cite{christensen_simulations_2025}.} reflection. This is done because the copper (311) reflection has a much larger secondary extinction effect than Germanium, and therefore provides a good benchmark against the transfer equations.

The files used for simulating these experiments can be found in \cite{christensen_simulations_2025}

\FloatBarrier
\section{Results}

\begin{figure}
    \centering
    \includegraphics[width=0.5\linewidth]{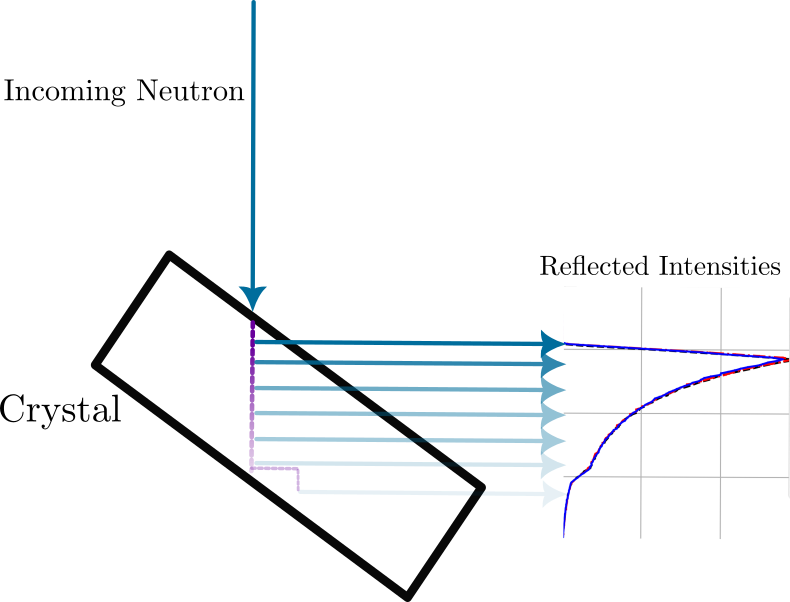}
    \caption{Visualization of the effect of attenuation throughout the unbent mosaic crystal. The purple dots in the crystal illustrate potential paths of the neutrons.}
    \label{fig:Explaining_attenuation}
\end{figure}

Two types of simulated measurements were carried out: (I) the spatial distribution of the reflected beam registered by the PSD monitor and (II) a rocking curve recorded by simulating the integral intensity at the monitor as a function of the rocking angle, $\omega$. In the first case (I), the result shows the distribution of the scattering power (modulated by absorption) along the neutron path, as illustrated in Figure \ref{fig:Explaining_attenuation}. The second case (II) elucidates the magnitude and angular width of the crystal reflectivity.

\label{sect:results}
\subsection{Bent perfect crystal}
\label{sec:Bent_perfect_crystal}
\begin{figure}
    \centering
    \includegraphics[width=\linewidth]{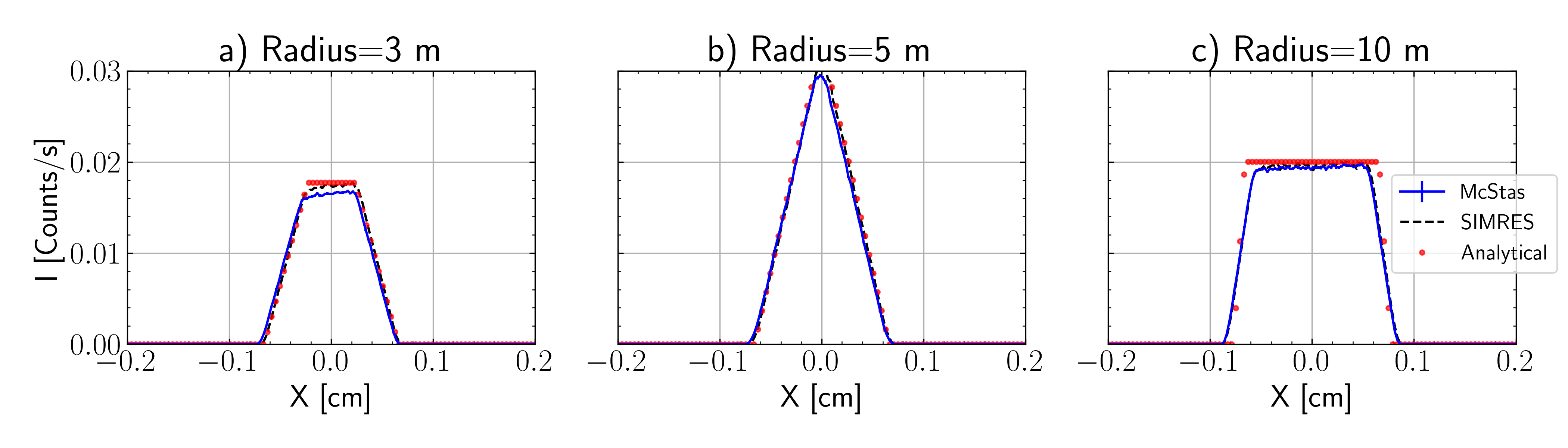}
    \caption{SIMRES and McStas simulations of the spatial distribution for Bragg scattering from a perfect crystal, bent at different radii. Overlain with the analytically calculated reflectivity.}
    \label{fig:X_profile_perfect}
\end{figure}

The simulated spatial distribution (I) of the reflected beam from the bent perfect crystal (case (i) $\eta = 0$) can be seen in Figure \ref{fig:X_profile_perfect}.  The results from McStas and SIMRES are almost identical, with the points of reflection localized near the crystal center as expected. The spatial distribution is determined by the contributions of primary beam width and angular divergence. For the monochromatic beam defined by a pair of distant slits of equal width, the resulting beam profile at the detector is a convolution of two uniformly distributed random contributions (see Appendix \ref{appendix:Bragg_Expansion}). This analytical solution is shown as the red dots on Figure \ref{fig:X_profile_perfect}.
For the bending radius near $R$=5 m, the widths of both contributions are equal, yielding the observed triangular beam profile.


\begin{figure}
    \centering
    \includegraphics[width=\linewidth]{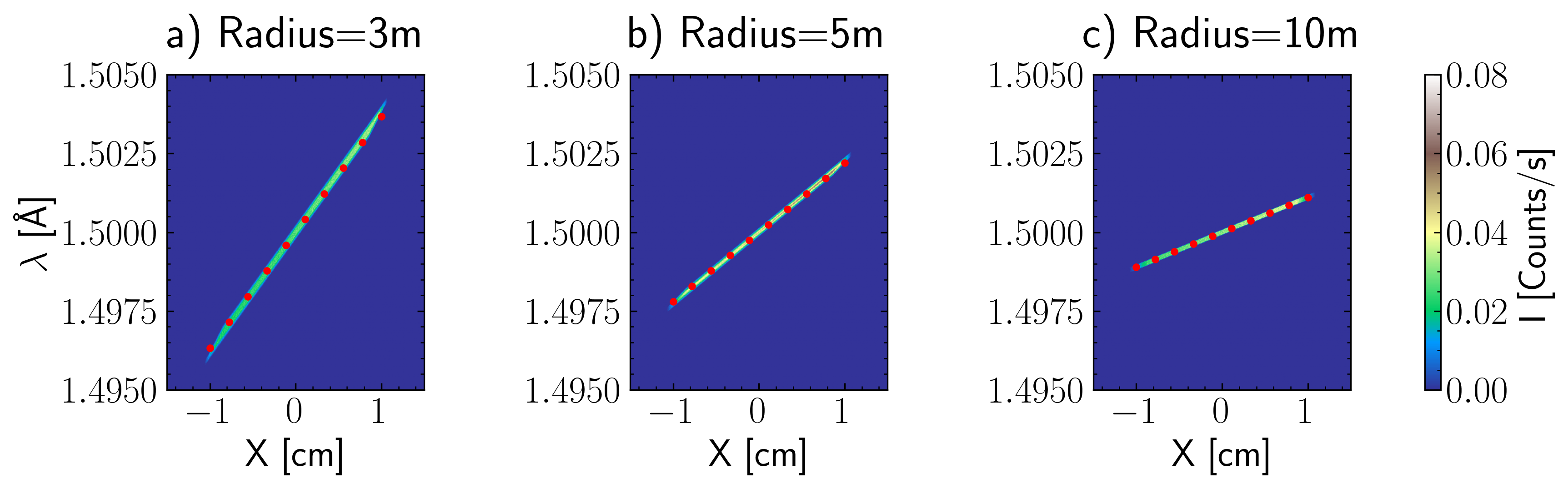}
    \caption{Two-dimensional plots of the correlations between wavelength and spatial position, corresponding to McStas simulations of crystal (i). Simulions are performed with a broad wavelength spectrum $\lambda = 1.5 \pm 1.5\cdot 10^{-2}$~\AA, as well as monitor which was 0.03~m $\times$ 0.02~m (width $\times$ height). The red dots lie with the theoretically expected slope, as calculated in the text.}
    \label{fig:spatial_wavelength_broad}
\end{figure}
To visualize the wavelength-space correlation, the simulation was repeated for a broader wavelength band when the effects of beam width and divergence can be neglected. The resulting Figure \ref{fig:spatial_wavelength_broad} illustrates how the position of reflection (projected on the detector plane) varies along the incident beam as a function of wavelength. As expected for uniform bending, the slope is linearly proportional to the curvature. Its value is determined by the condition $\epsilon_0 + \beta t = 0$ from Equation \ref{eq:eps_t}, where $\epsilon_0 = \tan(\theta) \Delta\lambda / \lambda $ is the deviation from Bragg condition for wavelength $\lambda + \Delta\lambda$. The time to reflection, $t$ then determines the position at the detector, $x_{det} = \mathbf{s}_D \cdot \mathbf{v}_i t$, where $\mathbf{s}_D$ is the unit vector along x-axis of the detector. This calculated dependence is indicated by the dots plot over the intensity maps in Figure \ref{fig:spatial_wavelength_broad}. 
In order to calculate this line, the $\beta$ value for the crystal with the specific $R$ was extracted from the simulation. These $\beta$ values are, $\beta_3=612$/s, $\beta_5=367$/s, $\beta_{10}=184$/s.

\begin{figure}
    \centering
    \includegraphics[width=\linewidth]{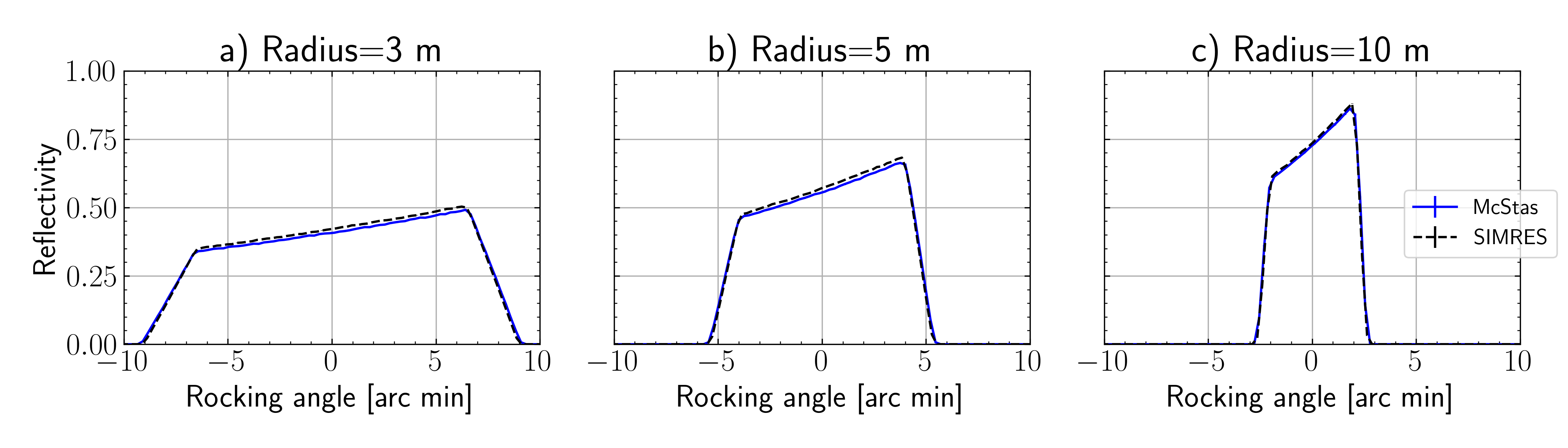}
    \caption{SIMRES and McStas simulations, showing rocking curves for a perfect crystal, bent at different radii.}
    \label{fig:Rocking_curve_perfect}
\end{figure}

The rocking curves (II) in Figure \ref{fig:Rocking_curve_perfect} become wider with increasing curvature $1/R$. This is reasonable, as the growing curvature increases the spread of lattice plane angles. The peaks are all asymmetric. This is caused by a change in penetration length, as the point of reflection moves within the crystal.

\subsection{Bent mosaic crystal}
\begin{figure}
    \centering
    \includegraphics[width=\linewidth]{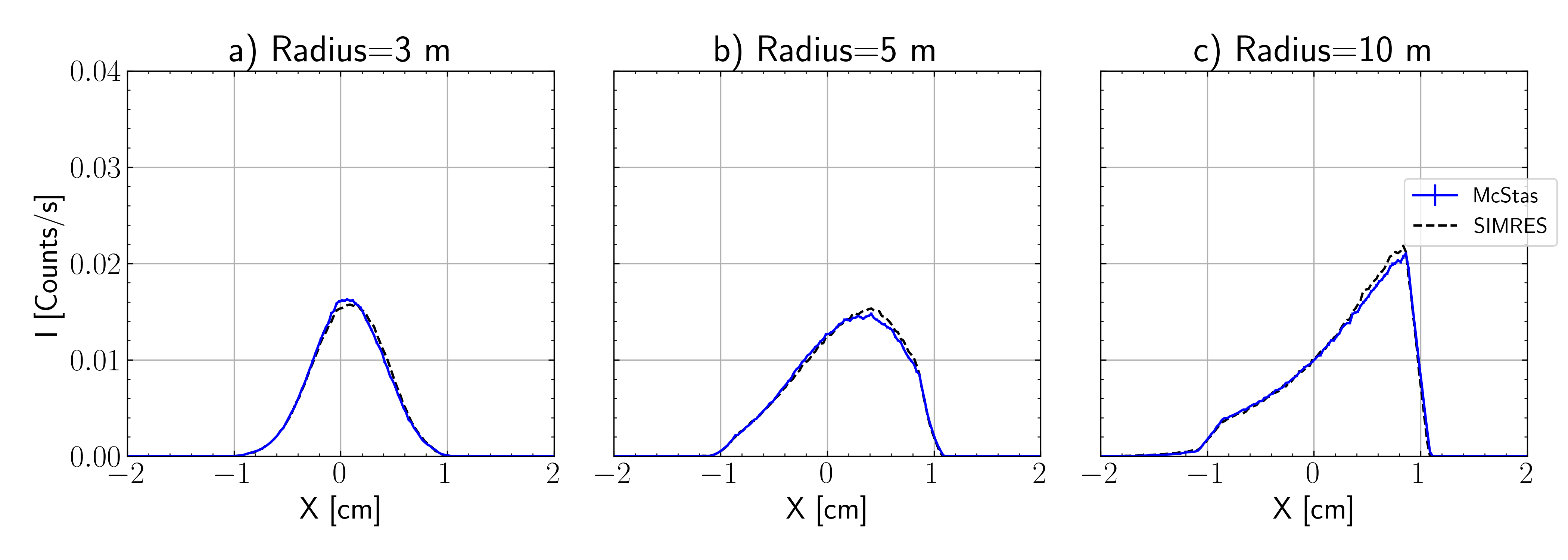}
    \caption{McStas and SIMRES simulations of the spatial distribution of the reflected neutrons. Simulations performed for the crystal of case (ii).}
    \label{fig:X_profile_mosaic}
\end{figure}

(I) The spatial distributions of the reflected neutrons for the bent mosaic crystal (case (ii)) can be seen in Figure \ref{fig:X_profile_mosaic}. This data illustrates the transition between the limiting cases from the bent perfect crystal ($\eta = 0$) and flat mosaic crystal ($1/R = 0$). For $R=3$ m, the broadening caused by mosaicity is small and the reflection points become localized around the crystal center. In the other end, at $R = 10$ m, the effect of mosaicity is dominant and the distribution approaches that of a flat mosaic crystal. 

\begin{figure}
    \centering
    \includegraphics[width=\linewidth]{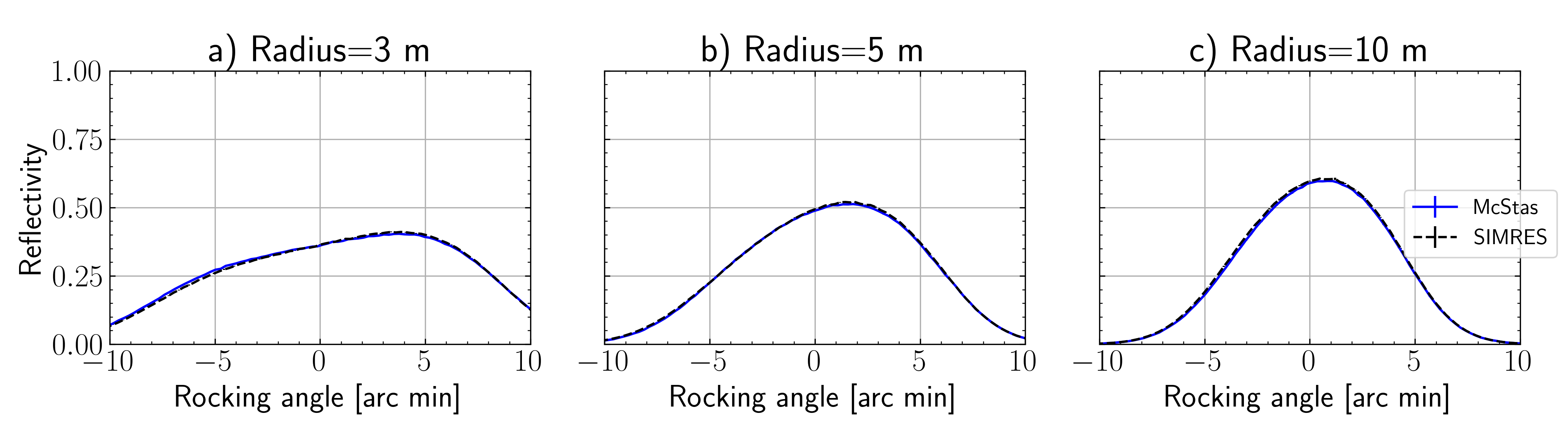}
    \caption{McStas and SIMRES simulations the crystal in case (ii). The simulations are rocking curves for a mosaic crystal, bent at different radii.}
    \label{fig:Rocking_curve_mosaic}
\end{figure}

(II) The rocking curves of the mosaic crystal bent at different radii are presented in Figure \ref{fig:Rocking_curve_mosaic}. They also illustrate the transition from weak bending ($R=10$ m) when the peak shape is determined mainly by the Gaussian mosaicity, to the case of strong bending ($R=3$ m) where the curve is a convolution of the crystal mosaicity distribution with the bent perfect crystal rocking curve shown in  \ref{fig:Rocking_curve_perfect}. 

\FloatBarrier
\subsection{Nearly unbent mosaic crystal}
In the case of negligible crystal curvature, it is possible to compare results with another McStas simulation using the NCrystal model for mosaic crystals. 

\begin{figure}
    \centering
    \includegraphics[width=\linewidth]{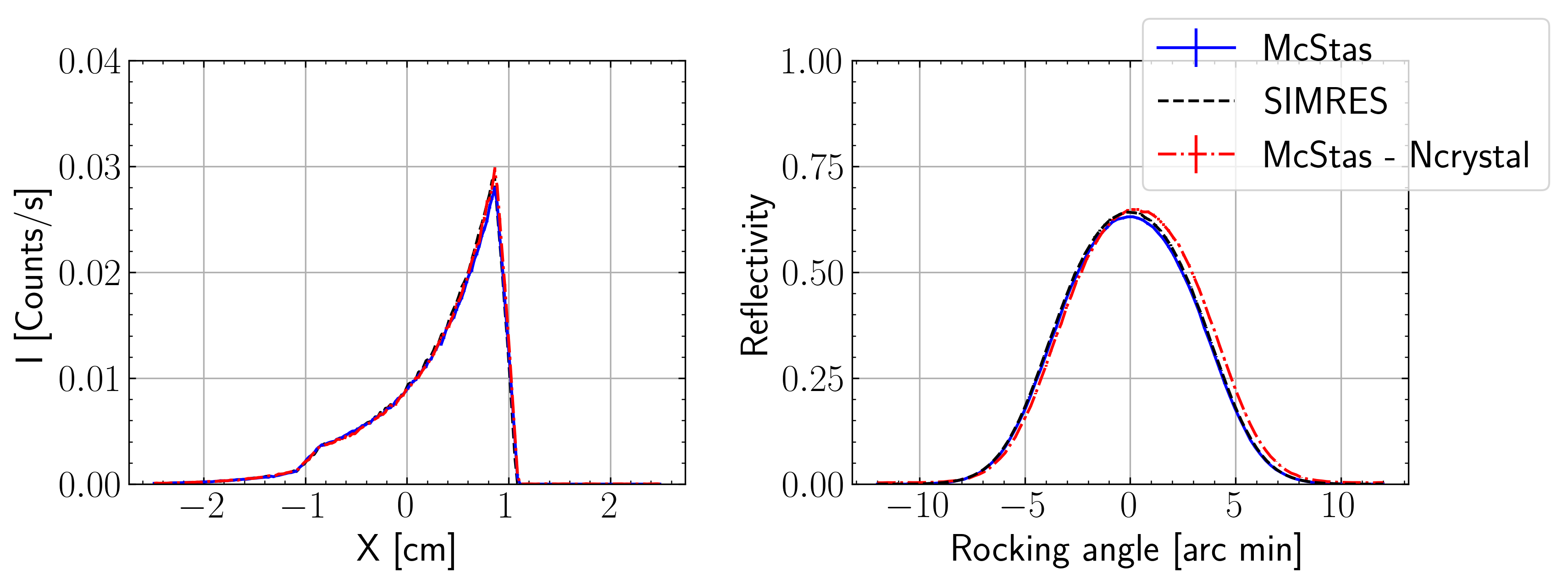}
    \caption{(Left) Spatial distribution and (Right) rocking curve for the (almost) flat mosaic crystal.}
    \label{fig:flat_crystal_space_rock}
\end{figure}

(I) For the spatial distribution of the almost unbent 6' mosaic crystal (case (iii)), Figure \ref{fig:flat_crystal_space_rock} (left) shows a gradual increase in intensity along $X$, corresponding to a shorter distance travelled through the crystal. This effect is due to the attenuation from absorption and extinction, as is visualized on Figure \ref{fig:Explaining_attenuation}.

(II) The rocking curve of the nearly unbent mosaic crystal is likewise presented in Figure \ref{fig:flat_crystal_space_rock} (right). This rocking curve is quite close to the gaussian mosaicity distribution as expected if the effect of multiple diffraction (secondary extinction) is weak.

The agreement with the NCrystal model is almost perfect and thus provides an additional validation of the new McStas component. However, the zero point of the rocking curves is different by a third of an arc minute, with our component being centered at zero.

\begin{figure}
    \centering
    \includegraphics[width=0.8  \linewidth]{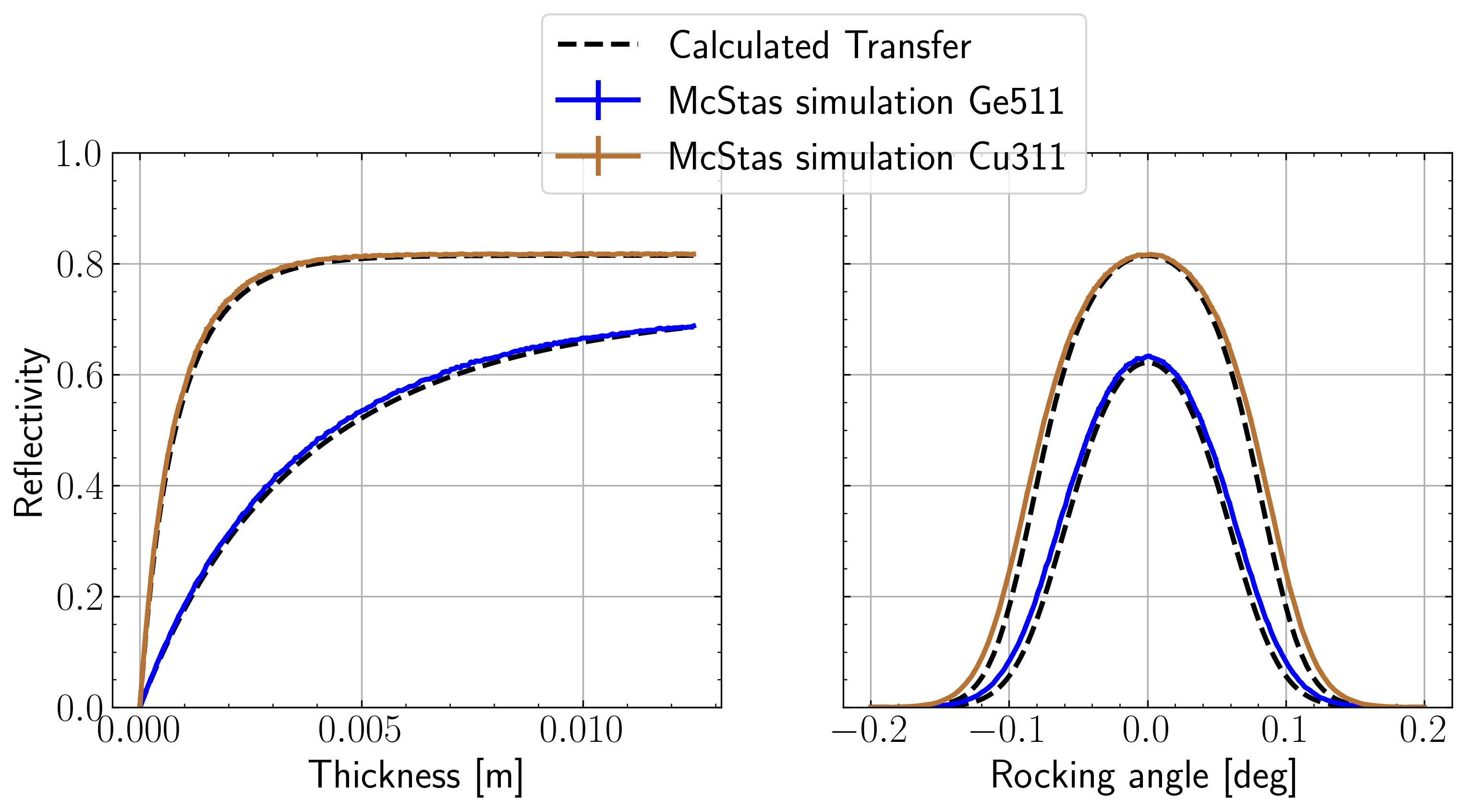}
    \caption{Comparison of analytical calculations and McStas simulations in the case (iii) of flat mosaic crystal at $\lambda = 1.5$~\AA. (Left) peak reflectivity (maxima of the rocking curves) as a function of crystal thickness. (Right) Rocking curves for the crystal thickness of 8 mm.}
    \label{fig:calculation_comparison}
\end{figure}

To further examine the effect of secondary extinction, we compare the simulation of the rocking curves for Ge(511) with that of Cu(311) for the same crystal and instrument geometries. The reflectivity for Cu(311) is significantly higher that of Ge(511) and rocking curve broadening is thus expected as a result of multiple diffraction. These simulations are also compared with the analytical solution to the Darwin intensity transfer equations \cite{lefmann_neutron_1997} given in Equation \ref{eq:Analytical_mosaic_reflectivity}. The results can be seen in Figure \ref{fig:calculation_comparison}. The calculations use the same values of kinematic reflectivity as the simulations, namely $Q_{Ge}=0.17$ m$^{-1}$ and $Q_{Cu}=1.14$ m$^{-1}$. These kinematical reflectivities, have been multiplied by the Debye Waller factor for the material, which is calculated in accordance with \cite{freund_cross-sections_1983}. The absorption coefficients used at this wavelength were $\mu_{Ge}=15$ m$^{-1}$ and $\mu_{Cu}=45$ m$^{-1}$, and the primary extinction factors were $\Upsilon_{Ge} = 0.95$, $\Upsilon_{Cu} =0.77 $.
A broad, non-gaussian curve shape is indeed observed for Cu(311) as a result of a strong secondary extinction effect.
 
The agreement between the simulations and the analytical calculations is good in this case of regular plane parallel crystal geometry, because the random walk algorithm in McStas effectively solves the intensity transfer equations for the same boundary conditions as those assumed in Equations \ref{eq:Analytical_mosaic_reflectivity}.

\FloatBarrier
\newpage 
\section{Discussion and conclusion}
\label{sect:conclusion}


In this work, we have developed a new McStas component \textit{Monochromator\_bent.comp} that simulates neutron transport in bent, and potentially mosaic, single crystals. Our work fills a gap for the McStas software package, in simulating monochromator and analyzer arrays, that use this type of crystals. Therefore this new development allows simulation of instrument designs that were previously impossible to simulate with McStas only. 

\textit{Monochromator\_bent} is based on the algorithm presented in \cite{NIMA}. Said algorithm is used due to its relative simplicity and therefore speed and reliability. In short, the algorithm calculates a probability for the neutron ray to reflect, and then chooses the distance the ray travels before reflection.
Multiple scattering is then performed by looping these two steps, potentially over an array of crystals. 

In order to validate our work, we performed simulations of two different experimental set-ups: 
(I) measurement of the spatial distribution of the reflected neutrons and (II) the reflected intensity in a rocking curve scan. These simulations were performed for three different extreme limits of the crystal. (i) a single crystal bent at three different bending radii $R=3,5,10$ m with $\eta=0'$. (ii) the same crystal, but with $\eta=6'$, and finally (iii) which was an almost flat single crystal $R=1000$ m, with $\eta=6'$. For the last case, a McStas simulation with NCrystal was also performed for independent comparison.
In addition, we simulated the reflectivity of a Ge(511) crystal and Cu(311) crystal as a function of thickness and rotation angle and compared the results to analytical calculations, in order to investigate if the effect of multiple scattering was as expected.

Case (i) showed that the reflected spatial distribution of the bent perfect crystal was a convolution of two uniform distributions, that is the width of the slits in the instrument setup. It also showed that the rocking curves broaden with increasing curvature.
In case (ii) we see that the the spatial distribution transitions from a signal modulated by the penetration length in the crystal, towards a signal localized around the crystal center, with increasing curvature $1/R$. The rocking curve likewise illustrates a transition from a gaussian curve due to the mosaicity at low curvature, to a convolution of the mosaicity distribution with the rocking curve distribution in case (i) of experiment (II).

The nearly unbent crystal (case (iii)) showed the limiting case where the crystal is almost an ideal type I crystal. Here, the spatial distribution almost only being modulated by absorption and single diffraction events. Here the rocking curve is almost purely gaussian for the Ge(511) crystal, indicating that multiple scattering effect 9s small. The simulations in this case were also compared to NCrystal, which achieved almost identical results. However, we notice that the rocking curve in the NCrystal case seem slightly shifted away from the center of the scan, although the widths of the rocking curve agrees with our new component. We attribute this difference to the skew of rocking curves shown by \citet{wuttke_multiple_2014}, to be present when solving the generalised Darwin-Hamilton equations. This asymmetry is also pointed out by the developers of the NCrystal software program in the treatment of their elastic modules \cite{kittelmann_elastic_2021}. In this special case, the results can also be compared with analytical calculations of reflectivity (Equation \ref{eq:Analytical_mosaic_reflectivity}), which yield almost equal rocking curves as the simulations.
Importantly this holds for the Cu(311) crystal, even though the rocking curve shape is distinctly non-gaussian.

What the simulations in all cases have in common is that SIMRES and \textit{Monochromator\_bent} are almost identical throughout all simulations, and very similar to what was earlier published in \cite{NIMA}.

Similarly to the previous work, \textit{Monochromator\_bent} relies on a number of approximations that limits its applicability: 
(A) The simulation model becomes invalid near the limit of bent perfect crystal ($R=0$, $\eta=0$, where the dynamical theory of diffraction has to be used.   
(B) The orientation distribution is gaussian and independent on position, which may not be a valid assumption for crystals produced by plastic bending. Comparison with experimental data would be needed the asses the model validity for various types of microstructure.
(C) Primary extinction is not simulated, but is approximately accounted for by multiplication of the $\Upsilon$ factor onto the kinematical reflectivity.
(D) Only a single lattice plane is assumed to contribute to diffraction, thereby not allowing the simulation of competing reflections. The beam attenuation at certain wavelengths may thus be inaccurate. 
(E) The bending applied is cylindrical, elastic and uniform, but not extreme. This means that this component should not be used to simulate crystals, where the bending radius is of the same order of magnitude as the crystal dimensions. 
(F) The equations derived do not consider reflecting planes which are not normal to the plane of bending (the $z-x$ plane in Figure \ref{fig:deformed_crystal_fig}). Hence the component \textit{Monochromator\_bent} does not support this case. 

In conclusion, the new McStas component \textit{Monochromator\_bent} simulates Bragg scattering in cylindrically bending single crystals under the assumptions stated above. The component has been compared to other verified algoritms from SIMRES and NCrystal, and the results are in almost perfect agreement. The component is currently being added to the McStas library.
Our component enables instrument scientists and neutron users to build McStas models that contain bent crystals. This will have impact both on design of future instrumentation, and for optimizations of current instruments. 

\section*{Acknowledgments}
DLC thanks the Institute Laue Langevin for hospitality during parts of this work, and ACTNXT for funding during parts of this work.

We thank Peter Willendrup and Mads Bertelsen from the ESS Data Management and Software Center for support with McStas.

JS acknowledges the supported by the Czech Ministry of Education, Youth and Sports (ESS Scandinavia-CZ, project LM2023057).

\newpage



\appendix

\section{Derivation of deformation gradient}
\label{appendix: deformation gradient}
The spatial dependence of $\boldsymbol{\tau}(\mathbf{r})$ can be derived from the displacement function $\zeta(z,x)$ which defines the neutral surface in a bent thin plate. The solution for small deformations has been found by \citet{POPOVICI} in the form 

\begin{align}
	\boldsymbol{\tau}(\mathbf{r})=
	\boldsymbol{\tau}_0\left(1+x\frac{\partial^2\zeta}{\partial z^2}(1-(1+\kappa)\cos^2(\chi))\right)
	&+ |\boldsymbol{\tau}_0|\mathbf{s}\left(\frac{\partial \zeta}{\partial z} - 
	x\frac{\partial^2\zeta}{\partial z^2}(1 + \kappa)\sin(\chi)\cos(\chi) \right)\nonumber \\
	& - |\boldsymbol{\tau}_0| \mathbf{s'}\left( \frac{\partial \zeta}{\partial y}\cos(\chi) + x\frac{\partial^2\zeta}{\partial z \partial y}\sin(\chi)\right).
	\label{eq: reciprocal lattice vector of bent crystal}
\end{align} 

where we use the notation described in Section \ref{sec_scattering_power} and Figure \ref{fig:deformed_crystal_fig}. The symbol $\kappa$ stands for the effective Poisson elastic constant, which depends on the type of boundary conditions, crystal elastic constants and lattice orientation. Values of $\kappa$ for particular cases relevant for monochromators and analyzers can be found in \cite{POPOVICI}). 

In this project, $\zeta$ is assumed to describe uniform elliptical bending $\zeta = \frac{z^2}{2R_h} + \frac{y^2}{2R_v}$. In many practical situations, the crystal bending is close to cylindrical and the vertical ($y$) component of curvature $1/R_v$ can be neglected. However, for very thin wafers or pneumatic bending, a non-zero vertical curvature has to be considered. 

The deformation gradient of $\boldsymbol{\tau}$ 
\begin{align}
	\nabla\boldsymbol{\tau} &= \begin{bmatrix}
		\frac{\partial\boldsymbol{\tau}_x}{\partial x} & \frac{\partial\boldsymbol{\tau}_x}{\partial y} &  \frac{\partial\boldsymbol{\tau}_x}{\partial z}\\
		\frac{\partial\boldsymbol{\tau}_y}{\partial x} & \frac{\partial\boldsymbol{\tau}_y}{\partial y} &  \frac{\partial\boldsymbol{\tau}_y}{\partial z}\\
		\frac{\partial\boldsymbol{\tau}_z}{\partial x} & \frac{\partial\boldsymbol{\tau}_z}{\partial y} &  \frac{\partial\boldsymbol{\tau}_z }{\partial z}
	\end{bmatrix}
\end{align}
is then derived by substitution for the direction vectors

First we express $\boldsymbol{\tau}_0$, $\boldsymbol{s}$ as well as $\boldsymbol{s}'$ as a function of $\chi$, and then we express $\boldsymbol{\tau}$ from \ref{eq: reciprocal lattice vector of bent crystal}:
\begin{align}
    \frac{\boldsymbol{\tau}_0}{|\boldsymbol{\tau}_0|} &= \begin{bmatrix}
        \cos(\chi)\\
        0\\
        \sin(\chi)
    \end{bmatrix} \qquad \qquad \qquad\qquad\qquad\qquad
    \boldsymbol{s} = \begin{bmatrix}
        \sin(\chi)\\
        0\\
        -\cos(\chi)
    \end{bmatrix} \qquad \qquad \qquad\qquad\qquad
    \boldsymbol{s}' = \begin{bmatrix}
        0\\
        1\\
        0
    \end{bmatrix}
\end{align}
in (\ref{eq: reciprocal lattice vector of bent crystal}) and evaluation of the derivatives:
\begin{align}
    \nabla\boldsymbol{\tau} &= |\boldsymbol{\tau}_0| \begin{bmatrix}
        -\kappa\frac{\cos(\chi)}{R_h} && 0 && \frac{\sin(\chi)}{R_h}\\
        0 && -\frac{\cos(\chi)}{R_v} && 0\\
        \frac{\sin(\chi)}{R_h} && 0 && -\frac{\cos(\chi)}{R_h}
    \end{bmatrix}
\end{align}

\section{Flight time to reflection}
\label{appendix:Bragg_Expansion}
In the case of bent perfect crystal ($\eta = 0$), the flight time to the reflection point can be solved analytically  by setting $\epsilon = \gamma = 0$ and $\mathbf{k}_i = \mathbf{k}_{0i} + \Delta\mathbf{k}$, where $\mathbf{k}_{0i}$ denotes the mean incident neutron wave vector. For a crystal perfectly aligned with given diffraction vector $\boldsymbol{\tau}_0$, the Bragg condition implies that $|\mathbf{k}_{0i} + \boldsymbol{\tau}_0| = |\mathbf{k}_{0i}|$ and expansion of Equation 5 from the main text then yields
																					  
\begin{align}
	2 \boldsymbol{\tau}_0 \cdot \Delta\mathbf{k} + 
	2 \left( \mathbf{k}_{0i} + \boldsymbol{\tau}_0\right) \cdot \boldsymbol{\nabla\tau} \cdot \mathbf{r} +
	2 \Delta\mathbf{k} \cdot \mathbf{r} +
	|\boldsymbol{\nabla\tau} \cdot \mathbf{r}|^2 = 0
\end{align}

By substituting for $\mathbf{r} = \mathbf{r}_0 + \mathbf{v}_i t$ and solving for $t$, we arrive at the quadratic equation for $t$,

\begin{align}
\begin{aligned}
	\epsilon_0 + 	
	2 \left(\mathbf{k}_{0i} + \boldsymbol{\tau}_0 + \boldsymbol{\nabla\tau} \cdot \mathbf{r}_0 + \Delta\mathbf{k}\right) 
	\cdot \left(\boldsymbol{\nabla\tau} \cdot \mathbf{v}_i\right) t + 
	|\boldsymbol{\nabla\tau} \cdot \mathbf{v}_i|^2 t^2  = 0 , \\
	\epsilon_0 \equiv 2 \boldsymbol{\tau}_0 \cdot \Delta\mathbf{k} +
    2 \left( \mathbf{k}_{0i} + \boldsymbol{\tau}_0 +\Delta\mathbf{k} \right) \cdot \boldsymbol{\nabla\tau} \cdot \mathbf{r}_0 + 
   |\boldsymbol{\nabla\tau} \cdot \mathbf{r}_0|^2 .	
\end{aligned} 
\end{align} 
In almost all practical situations, the 2nd order terms with $\boldsymbol{\nabla\tau}$ and $\Delta\mathbf{k}$ can be neglected, which leads to the linear solution

\begin{align}
	t = -\frac { \boldsymbol{\tau}_0 \cdot \Delta\mathbf{k} + 
		\left(\mathbf{k}_{0i} + \boldsymbol{\tau}_0\right) \cdot \boldsymbol{\nabla\tau} \cdot \mathbf{r}_0}
	{\left(\mathbf{k}_{0i} + \boldsymbol{\tau}_0\right) \cdot \boldsymbol{\nabla\tau} \cdot \mathbf{v}_i} . \label{eq:time_to_reflection}	
\end{align}  

For the simple instrument setup considered in this article, the incident neutron trajectory is determined by two random variables, which are the neutron positions at the pair of distant slits. Their distribution is uniform and can be expressed by a pair of uniform random numbers $\xi_0$ and $\xi_1$ on the interval (-0.5;0.5). We can thus substitute for $\mathbf{r}_0$ and $\Delta\mathbf{k}$:

\begin{align}
\boldsymbol{\mathbf{r}}_0 = \xi_1 w \mathbf{u}_T,  \qquad \qquad
\Delta\mathbf{k} = (\xi_1 - \xi_0) w/L \mathbf{u}_T ,  \qquad \qquad
\mathbf{u}_T \equiv
\begin{bmatrix}
	\cos(\theta + \chi)\\
	0\\
	\sin(\theta + \chi)
\end{bmatrix}
\end{align} 

where $w$ is the slit width, $L$ is their distance and $\mathbf{u}_T$ is the unit vector normal to the incident beam.
Substitution in \ref{eq:time_to_reflection} then allows to calculate the distribution of $t$, and hence that of the reflection positions, $\mathbf{r}_0 + \mathbf{v}_i t$ as a convolution of two uniform distributions. This results in the trapezoidal intensity distribution at the detector as observed in the simulations (see Section \ref{sec:Bent_perfect_crystal}). 



\section*{References}

\bibliography{bib}







\end{document}

%% file: Figures/Unbent_crystal.tex
\begin{tikzpicture}
    \def\scatPointX{1.2cm}
    \def\scatPointY{0.9cm}
    \def\boxTop{1.5cm}
    \def\boxBottom{-0.5cm}
    \def\boxCenter{0.5cm}

    \draw (-3.05,-0.5) -- (-3.05, -0.5)node[black, below right]{}; 
    \draw[dash pattern={on 2pt off 0.5pt}, semitransparent] (-3.05,\boxBottom) rectangle (3.05, \boxTop);
    \draw[dash pattern={on 2pt off 0.5pt}, semitransparent,] (-3.05,\boxCenter) -- (3.05,\boxCenter);

    \foreach \x in {-2.4, -1.8, ..., 2.4} {
        \draw[dash pattern={on 2pt off 0.5pt}, black, thin, semitransparent] (\x, \boxTop) -- (\x, \boxBottom);
    }
    \usetikzlibrary{calc,patterns,angles,quotes}
    \draw[->] (0,\boxCenter) node[below left] {$\boldsymbol{y}$} -- (3.5,\boxCenter) node[right] (z) {$\boldsymbol{z}$};
    \draw[->] (0,\boxCenter) -- (0,3.5) node[above] {$\boldsymbol{x}$};
    
    \draw[thick, ->] (0,\boxCenter) -- (\scatPointX-0.2cm,\scatPointY) node[midway, above] {$\boldsymbol{r}$};

    \begin{scope}[shift={(\scatPointX, \scatPointY)},rotate=-25]

        \draw[thick, ->] (0,0) -- (0,3cm) node[right] {$\boldsymbol{\tau_0}$};
        \draw[thick, ->] (2cm,0) -- (-2cm,0) node[above] {$\boldsymbol{s}$};
        \draw[thick] (-2cm,0) -- (3cm,0) node[above] (s) {};

    \end{scope}
    \coordinate (crossing) at (2.8cm,\boxCenter-0.2cm);
    
    \pic [draw, <->, "$\chi$", angle eccentricity=1.5] {angle = s--crossing--z};

    \node[overlay, draw, circle, fill=white, minimum size=0.01cm, text opacity=0] at (\scatPointX,\scatPointY) {};
    \node[overlay, scale=1.5] at (\scatPointX,\scatPointY) {$\otimes$};
    \node[overlay,] at (\scatPointX-0.1cm,\scatPointY + 0.4cm) {$\boldsymbol{s}'$};
\end{tikzpicture}

%% file: Figures/Single_crystal.tex
\begin{tikzpicture}

    \def\outerradius{4cm}  
    \def\innerradius{3cm}  
    \def\gridspacing{0.5cm} 
    \def\rotation{0}    
    \def\startAngle{220}
    \def\endAngle{320}
    \def\monoxshift{0cm}
    \def\monoyshift{3.5cm}
    \def\shiftx{0cm}
    \def\shifty{0cm}
    \definecolor{orange}{HTML}{C65102}

    \draw (-3.05,-0.5) -- (-3.05, -0.5)node[black, below right]{Unbent crystal}; 
    \draw[dash pattern={on 2pt off 0.5pt}, semitransparent] (-3.05,-0.5) rectangle (3.05, 0.5);
    \draw[dash pattern={on 2pt off 0.5pt}, semitransparent,] (-3.05,0) -- (3.05,0);

    \foreach \x in {-2.4, -1.8, ..., 2.4} {
        \draw[dash pattern={on 2pt off 0.5pt}, black, thin, semitransparent] (\x, 0.5) -- (\x, -0.5);
    }
    
    \draw[thick, xshift=\monoxshift, yshift=\monoyshift] (\startAngle:\outerradius) arc[start angle=\startAngle, end angle=\endAngle, radius=\outerradius];

    \draw[thick, xshift=\monoxshift, yshift=\monoyshift] (\startAngle:\innerradius) arc[start angle=\startAngle, end angle=\endAngle, radius=\innerradius];

    \draw[thick, xshift=\monoxshift, yshift=\monoyshift] (\startAngle:\outerradius) -- (\startAngle:\innerradius)node[above left]{Bent crystal};
    \draw[thick, xshift=\monoxshift, yshift=\monoyshift] (\endAngle:\outerradius) -- (\endAngle:\innerradius);

    \begin{scope}[xshift=\monoxshift, yshift=\monoyshift]
        \clip (\startAngle:\outerradius) arc[start angle=\startAngle, end angle=\endAngle, radius=\outerradius] -- (\endAngle:\innerradius) arc[start angle=\endAngle, end angle=\startAngle, radius=\innerradius] -- cycle;
        
        \begin{scope}[rotate=0]
            \draw[gray, thin, xshift=\monoxshift, yshift=0] (\startAngle:3.5cm) arc[start angle=\startAngle, end angle=\endAngle, radius=3.5cm];

            \foreach \x in {\startAngle, 230, 240, ..., \endAngle} {
                \draw[gray, thin] (\x:\outerradius) -- (\x:\innerradius);
            }
        \end{scope}
    \end{scope}

    \draw[->] (0,0) node[below left] {$y$} -- (3.5,0) node[right] {$\boldsymbol{z}$};
    \draw[->] (0,0) -- (0,3.5) node[above] {$\boldsymbol{x}$};

    \draw[thick, ->] (0,0) -- (2.05cm,0.65cm) node[right] {$\boldsymbol{r}$};

    \draw[thick, ->] (2.05cm,0.65) -- (2.05cm,3cm) node[right] {$\boldsymbol{\tau_0}$};
    \draw[thick, ->] (2.05cm,0.65) -- (0.8cm,2.5cm) node[right] {$\boldsymbol{\tau}(\boldsymbol{r})$};
\end{tikzpicture}

%% file: Figures/instrument_setup.tex
\begin{tikzpicture}[scale=0.8]
    \filldraw[color=black!60, fill=red!0, very thick](0,-3) circle (1) node[color=black]{Source};

    \draw[color=neutron,ultra thick] (0,-3.4) -- (0,-10.3);
    \draw[] (0,-8) -- (0,-8)node[color=neutron, above right]{$\boldsymbol{k}_i$};
    \draw[color=neutron,ultra thick, ->] (0,-10.3) -- (7,-7);
    \draw[] (4,-8.2) -- (4,-8.2)node[color=neutron, above right]{$\boldsymbol{k}_f$};

    \draw[color=black, thick] (-1,-5) node[color=black, above left]{Slit}-- (-0.3,-5);
    \draw[color=black, thick] (1,-5)-- (0.3,-5);

    \draw[color=black, dashed] (-1,-5.5) -- (1,-5.5) node[color=black, below right]{Normalizing monitor};

    \draw[color=black, dashed, xshift=0.38cm] (7.2,-8) -- (6,-6) node[color=black, above right]{PSD Monitor};
    \node at (7.3,-6.85) [color=black, rotate=120] {X};

    \filldraw[color=black!0, very thick](0,-10.3) circle (0.12) node[color=black]{$\bigotimes$};
    \draw[color=black, thick, ->] (0,-10.3)node[color=black, below left]{$\boldsymbol{y}$} -- (2.5,-12.7)node[color=black, above right]{$\boldsymbol{z}$};
    \draw[color=black, thick, ->] (0,-10.3) -- (3,-7.5)node[color=black, above right]{$\boldsymbol{x}$};

    \def\outerradius{4cm}  
    \def\innerradius{3cm}  
    \def\gridspacing{0.5cm} 
    \def\rotation{-45}    
    \def\startAngle{220}
    \def\endAngle{320}
    \def\monoxshift{7cm}
    \def\monoyshift{-4cm}
    \def\shiftx{0cm}
    \def\shifty{0cm}

    \begin{scope}[rotate=\rotation]
        \draw[thick, xshift=\monoxshift, yshift=\monoyshift] (\startAngle:\outerradius) arc[start angle=\startAngle, end angle=\endAngle, radius=\outerradius] ;

        \draw[thick, xshift=\monoxshift, yshift=\monoyshift] (\startAngle:\innerradius) arc[start angle=\startAngle, end angle=\endAngle, radius=\innerradius];
    
        \draw[thick, xshift=\monoxshift, yshift=\monoyshift] (\startAngle:\outerradius) -- (\startAngle:\innerradius);
        \draw[thick, xshift=\monoxshift, yshift=\monoyshift] (\endAngle:\outerradius) -- (\endAngle:\innerradius);
        
        \begin{scope}[xshift=\monoxshift, yshift=\monoyshift]
            \clip (\startAngle:\outerradius) arc[start angle=\startAngle, end angle=\endAngle, radius=\outerradius] -- (\endAngle:\innerradius) arc[start angle=\endAngle, end angle=\startAngle, radius=\innerradius] -- cycle;
            
            \draw[gray, thin, xshift=\monoxshift, yshift=0] (\startAngle:3.5cm) arc[start angle=\startAngle, end angle=\endAngle, radius=3.5cm];

            \foreach \x in {\startAngle, 230, 240, ..., \endAngle} {
                \draw[gray, thin] (\x:\outerradius) -- (\x:\innerradius);
            }
        \end{scope}    
    \end{scope}
\draw[<->, thick, black, postaction={decorate}, 
    decoration={markings, mark=at position 0.5 with {\node[xshift=0.3cm]{$\omega$};}},
    xshift=\monoxshift, yshift=\monoyshift]
    (-4.1,-7.9) arc[start angle=355, end angle=375, radius=3.6cm];

\draw[<->, thick, black, xshift=\monoxshift, yshift=\monoyshift] 
    (-7.7,-3.2) arc[start angle=72, end angle=95, radius=3.2cm];

\end{tikzpicture}